\documentstyle[12pt,aps,prd,preprint,graphicx]{revtex}
\begin{document}
\title{Late-Time Evolution of Charged Gravitational Collapse and Decay of
Charged Scalar Hair - III. Nonlinear Analysis}
\author{Shahar Hod and Tsvi Piran}
\address{The Racah Institute for Physics, The
Hebrew University, Jerusalem 91904, Israel}
\date{\today}
\maketitle

\begin{abstract}
We study the {\it nonlinear} gravitational
collapse of a {\it charged} massless scalar-field.
We confirm the existence of oscillatory 
inverse power-law tails along future timelike
infinity, future null infinity and along the future outer-horizon.
The {\it nonlinear} dumping exponents are in excellent agreement with the 
{\it analytically} predicted ones.
Our results prove the analytic conjecture according to which
a {\it charged} hair decays {\it slower} than a neutral one and
also suggest the occurrence of mass-inflation along the Cauchy horizon
of a {\it dynamically} formed charged black-hole.
\end{abstract}

\section{introduction}\label{introduction}

Linearized {\it perturbation} analysis have revealed important
dynamical features of the gravitational collapse of massless neutral fields.  
In particular, according to the linearized perturbation theory there
are two major features which characterize the {\it late}
stages of the evolution: quasinormal ringing and inverse power-law
tails (which follows the quasinormal ringing).
These late-time tails are relevant to two major
aspects of black-hole physics: The {\it no-hair theorem} of Wheeler
and the {\it mass-inflation} scenario \cite{Poisson}.
The mechanism responsible for the development of these neutral inverse
power-law tails was first studied by Price \cite{Price}.
This work was further extended by Gundlach, Price and
Pullin \cite{Gundlach1}.
These authors have also shown
that the nonlinear dumping exponents, describing the fall-off of a
massless neutral scalar-field at late-times, are in good agreement with the prediction of the
linearized perturbation analysis \cite{Gundlach2}.

The late-time evolution of a {\it charged} massless 
scalar-field and     
the physical mechanism for the radiation of a {\it charged} 
hair were first studied {\it analytically} 
in \cite{HodPir1,HodPir2},
hereafter referred to as papers I and II, respectively.
The main result presented there is the existence of oscillatory 
inverse power-law tails along the asymptotic regions of future timelike
infinity, future null infinity and along the future outer-horizon with
{\it smaller} (compared to the
neutral case) dumping exponents.
These dumping exponents depend on the black-hole's (or the star's) charge.
In this paper we study numerically the 
fully {\it nonlinear} gravitational
collapse of a massless {\it charged} scalar-field.
We confirm that the late-time behaviour of the fully {\it nonlinear}
evolution is in excellent agreement with the {\it analytical}
predictions of the linearized analysis.  

The plan of the paper is as follows. In Sec. \ref{Sec2} we give a short
review of the linearized {\it analytical} results of papers I and II,
on which we base our expectations for the fully nonlinear case.
In Sec. \ref{Sec3} we describe our physical system, namely
the coupled Einstein-Maxwell-charged scalar equations.
In Sec. \ref{Sec4} we study the late-time evolution of the
{\it nonlinear} spherical charged gravitational collapse.
In Sec. \ref{Sec5} we study the late-time evolution
of non-spherical perturbations on a fixed 
Reissner-Nordstr\"om background and on a time-dependent background.
In all cases we compare the {\it nonlinear} results with the
predictions of the linearized {\it analytical} theory.
We conclude in Sec. \ref{Sec6} with a brief summary of our results and
their physical implications.

\section{Review of Linearized Analytical Results}\label{Sec2}

We study numerically the late-time behaviour of the 
fully {\it nonlinear} gravitational
collapse of a massless {\it charged} scalar-field. Our expectations
are based on the {\it linearized analytical} results of papers I and
II. Due to the {\it smallness} of the field's amplitude at late times,
we expect these results to hold even in the fully {\it nonlinear} case. In
other wards, we expect that the late-time field may be regarded as 
a small perturbation.
Thus, our expectations include the existence of inverse {\it power-law}
tails along the three asymptotic
regions: timelike infinity $i_+$, future null infinity $scri_+$ and along
the black-hole outer-horizon $H_{+}$ (where the power-law is multiplied by a 
{\it periodic} term).
Quantitatively, in paper II it was found that the {\it late-time}
behaviour of linearized {\it charged}-scalar perturbations (on a 
Reissner-Nordstr\"om background) is dominated
by power-law tails of the form
\begin{equation}\label{Eq1}
\psi^{l}_{m} =A_{ti}(l,eQ) y^{\beta +1} t^{-(2\beta +2)}\  , 
\end{equation}
at timelike infinity $i_{+}$,
\begin{equation}\label{Eq2}
\psi^{l}_{m} =A_{ni}(l,eQ) v_{e}^{-ieQ} u_{e}^{-(\beta +1-ieQ)}\  ,
\end{equation}
at future null infinity $scri _{+}$ and
\begin{equation}\label{Eq3}
\psi^{l}_{m} =A_{eh}(l,eQ) e^{i {{eQ} \over {r_{+}}} y} v_{e}^{-(2\beta +2)}\  ,
\end{equation}
along the black-hole outer-horizon $H_{+}$, where the charged
scalar-field $\phi$ is given by
\begin{equation}\label{Eq4}
\phi
=e^{ie\Phi t}\sum\limits_{l,m} {\psi _m^l\left( {t,r} \right)Y_l^m{{\left( {
\theta ,\varphi} \right)} \mathord{\left/ {\vphantom {{\left( {  \theta ,\varphi
 } \right)} r}}\right.}r}}\  ,
\end{equation}
where $\Phi$ is merely
a gauge constant of the electromagnetic potential $A_t$, i.e. its value at
infinity ($A_{t}$ is given by $A_{t}=\Phi -{Q \over r}$).
The constant $\beta (l,eQ)$ is given by
\begin{equation}\label{Eq5}
\beta ={{-1+ \sqrt {(2l+1)^{2} -4(eQ)^{2}}} \over 2}\  .
\end{equation}
The tortoise radial coordinate $y$ is defined by $dy=dr/\lambda^2$, where
$\lambda^2={1-{{2M} \over r}+{{Q^2} \over {r^2}}}$.
Here $v_{e} \equiv t+y$ and $u_{e} \equiv t-y$ are
the ingoing and outgoing Eddington-Finkelstein null coordinates, respectively.
The expressions for the coefficients $A(l,eQ)$ are given in paper II.

\section{The Einstein-Maxwell-charged scalar Equations}\label{Sec3}

We consider a spherically-symmetric self-gravitating
{\it charged} scalar-field $\phi$. The system is described by
the coupled Einstein-Maxwell-charged scalar equations.
This physical system was already described in a previous paper
\cite{HodPirm1}. Here we give only a short description of
the system and the final form of the equations studied.
We express the metric of a spherically symmetric spacetime in the
form \cite{Christodoulou,GoldPir}
\begin{equation}\label{Eq6}
ds^{2}=-g(u,r) \bar g(u,r) du^{2} -2g(u,r)dudr +r^{2}d\Omega ^{2}\  ,
\end{equation}
in which $u$ is a retarded time null coordinate and the
radial coordinate $r$ is a geometric quantity which directly measures
surface area.
The coordinates have been normalized so that $u$ is the 
proper time on the $r=0$ central world line.
We use the auxiliary field $\tilde \phi$ defined by
\begin{equation}\label{Eq7}
\phi \equiv {1 \over r} \int_{0}^{r} \tilde \phi \ dr\  ,
\end{equation}
in terms of which the Einstein equations yield
\begin{equation}\label{Eq8}
g(u,r)=exp \left [4 \pi \int_{0}^{r} {{(\tilde \phi- \phi)(\tilde \phi^{*}-
\phi^{*})}
\over {r}} dr \right]\  ,
\end{equation}    
and
\begin{equation}\label{Eq9}
\bar g(u,r)={1 \over r} \int_{0}^{r} \left (1- {Q^{2} \over {r^{2}}}
\right ) g\ dr\  .
\end{equation}
The electromagnetic potential $A_{u}$ is given
by the Maxwell equations
\begin{equation}\label{Eq10}
A_{u}=\int_{0}^{r} {Q \over {r^{2}}} g\ dr\  ,
\end{equation}
where the charge $Q(u,r)$ within a sphere of
radius $r$, at a retarded time $u$ is given by
\begin{equation}\label{Eq11}
Q(u,r)=4\pi ie \int_{0}^{r} r(\phi^{*} \tilde \phi- \phi \tilde \phi^{*}) dr\  .
\end{equation}
The mass $M(u,r)$ within a sphere of
radius $r$, at a retarded time $u$ is given by
\begin{equation}\label{Eq12}
M(u,r)= \int_{0}^{r} \left [2\pi {\bar g \over g}
(\tilde \phi- \phi)(\tilde \phi^{*}-\phi^{*}) +{1 \over 2} 
{{Q^{2}} \over {r^{2}}} \right ] dr +{1 \over 2} {Q^{2} \over r}\  .
\end{equation}
Finally, the 
wave-equation for the charged scalar-field 
takes the form of a pair of coupled differential equations
\begin{equation}\label{Eq13}
{d \tilde \phi \over du}={1 \over {2r}}(g - \bar g)(\tilde \phi- \phi)-
{{Q^{2}} \over {2r^{3}}}(\tilde \phi- \phi)g -{ieQ \over 2r}g \phi -
ie \tilde \phi A_{u}\  ,
\end{equation}
and
\begin{equation}\label{Eq14}
{dr \over du}={-{1 \over 2}} \bar g\  .
\end{equation}
We solve this system of equations numerically.
A detailed description of our algorithm, numerical methods, discretization
and error analysis are given in Ref. \cite{HodPirm1}.

In order to compare our {\it nonlinear} results with the linearized
{\it analytical} results of papers I and II, we use another
spacetime coordinate, namely the Bondi time $t_{B} \equiv t(u,\infty)$, which 
is the retarded time coordinate that agrees with time at
infinity. The proper time $t(u,r)$ along an $r=const$
trajectory is given by \cite{Gundlach2}
\begin{equation}\label{Eq15}
t(u,r) \equiv \int_{0}^{u} \sqrt{g(u',r) \bar g(u',r)} du'\  .  
\end{equation}
The comparison with the analytical results also requires the usage of
the ingoing and outgoing Eddington-Finkelstein null coordinates
$v_{e}$ and $u_{e}$.
In the asymptotic region $r \gg M_{BH}$, $v_{e}$ 
is linear with $r$ (along a $u=const$ ray). In particular, $v_{e} \simeq 2y \simeq 2r$
along the $u=u_{e}=0$ ray.
In a similar way $u_{e}$
is linear with $t$ along a $v=const$ ray, namely $u_{e}=2t-v_{e}$ 
(one may also use the asymptotic
relation $u_{e} \simeq v_{e}-2r$ along a $v=const$ ray in order to
evaluate $u_{e}$).
The relation between $A_{u}$ and $A_{t}$ implies that
$A_{t}(r_{+})=0$, i.e. $\Phi={Q \over {r_{+}}}$.

\section{Nonlinear Spherical Collapse}\label{Sec4}

In this section we present our results for 
the {\it nonlinear} spherically symmetric gravitational collapse 
of the self-gravitating massless {\it charged} scalar-field.
We have focused our attention on the behaviour of the charged field $\phi$
along the three
asymptotic regions: timelike infinity $i_{+}$, null infinity $scri_{+}$
and the black-hole outer horizon $H_{+}$.
The late-time evolution of a charged scalar-field is independent of
the form of the initial data. The numerical results presented
here have an initial profile of the form
\begin{equation}\label{Eq16}  
\phi(u=0,r)=Ar^{2}exp \left \{-\left [ (r-r_{0})/ \sigma \right ]^{2}
\right \}\  ,
\end{equation}
where $r_{0}=2.5, \sigma =0.5$ for the real part of the complex field
and  $r_{0}=3.0, \sigma =0.5$ for the imaginary part.
Figure 1 displays the time evolution of $\phi_{R}$, the real part of 
the charged scalar-field, along these asymptotic regions for a 
{\it supercritical} evolution ($A=0.005, e=0.85$) in which
a black-hole forms.
The mass and charge of the formed black-hole are $M_{BH}=0.503$ and
$Q_{BH}=-0.420$, respectively.
The top panel shows the behaviour of the field at a constant radius
(here $r=10$) as a function of the Bondi time.
The middle panel shows the behaviour of the field
along null infinity (approximated by the null surface $v=v_{max}$, where $v_{max}$
is the largest value of $v$ on the grid) as a function of $u_{e}$.
The bottom panel shows the behaviour of the field along the
black-hole outer horizon $H_{+}$ (approximated by the null surface
$u=u_{max}$, where $u_{max}$
is the largest value of $u$ on the grid) as a function of $v_{e}$.
Initially, the
evolution is dominated by the prompt contribution and by the quasinormal
ringing. However, at late-times a definite oscillatory {\it power-law}
fall off is manifest. The {\it nonlinear} power-law exponents (determined
from the maximas of these oscillations) are $-1.78$ at timelike infinity,
$-0.86$ along null infinity and $-1.96$ along the black-hole outer horizon.
These values should be compared with the {\it analytically} 
predicted values of $-1.70,-0.85,-1.70$, [see 
Eqs. ($\ref{Eq1}) -(\ref{Eq3})$], respectively. 
The theoretical values of the oscillation frequencies of the charged scalar field $\phi$ are
${{eQ_{BH}} \over {r_{+}}}$ at timelike infinity and along the
black-hole outer horizon and ${{eQ_{BH}} \over {2r_{+}}}$
along null infinity [see Eq. (\ref{Eq4}) and the 
relation $\Phi={Q \over r_{+}}$].
These values agree to within $2\%$ with the numerical estimates.

While the dumping exponents and the decay-rate of neutral
perturbations are independent of the spacetime parameters ($M$ and
$Q$), the analysis of paper II predicts that the {\it charged}
dumping exponents depend on the black-hole's (or the star's) 
{\it charge} (namely, on the dimensionless quantity $|eQ|$).
In order to test this linearized {\it analytical} prediction, we have 
studied the dependence of the charged dumping exponents at timelike
infinity on the 
parameter $e$. The initial form of the field is given by (16)
and the amplitude is set to $A=0.005$.
The nonlinear results are summarized in table \ref{Tab1}.
We find a good agreement between the numerically measured exponents
and the linearized predicted ones. 

The analysis of paper I predicts the existence of charged 
power-law tails even in subcritical evolution when there is
no collapse to a black-hole (and all we have are imploding 
and exploding shells). This phenomena is related
to the fact that the late-time evolution of the field is
dominated by the backscattering from asymptotically {\it far}
regions, and it does {\it not} depend on the small-$r$ details
of the spacetime (this is also the situation for neutral
massless perturbations \cite{Gundlach1}).
Since the charge of the spacetime is a {\it dynamical} quantity
it is not clear which value
of $Q$ should be taken in calculating the value of the dumping
exponent. However, since the charge of the spacetime falls to zero asymptotically
we expect the effective value of $|eQ|$ to be much {\it smaller} than unity.
In this case the dumping exponents are expected to be $2l+2$
at timelike infinity and $l+1$ along null 
infinity (with small corrections of order $O[(eQ)^{2}]$).
Figure 2 displays the time evolution of $\phi_{R}$ for a 
{\it subcritical} evolution ($A$=0.002, $e$=3).
Shown are the behaviour of the field at a constant radius
(here $r=10$ ) as a function of Bondi time and along null infinity 
as a function of $u_{e}$.
The {\it nonlinear} power-law exponents
are $-2.00$ at timelike infinity
and $-1.00$ along null infinity.
These values are exactly equal to those predicted in paper I.

\section{Non-Spherical Perturbations of Spherical Collapse}\label{Sec5}

In order to study the dependence of the late-time dumping-exponents
on the multipole index $l$ we have performed two series of 
numerical investigations.
First, we have integrated the {\it linearized charged} scalar-field equation
\begin{equation}\label{Eq17}
\psi _{,tt}+2ie{Q \over r}\psi _{,t}-\psi _{,yy}+V\psi =0\  ,
\end{equation}
where
\begin{equation}\label{Eq18}
V=V_{M,Q,l,e}\left( r \right)=\left( {1-{{2M} \over r}+{{Q^2} \over {r^2}}}
\right)\left[ {{{l\left( {l+1} \right)} \over {r^2}}+{{2M} \over {r^3}}-{{2Q^2}
\over {r^4}}} \right]-e^2{{Q^2} \over {r^2}}\  ,
\end{equation}
on a {\it fixed} Reissner-Nordstr\"om background.
It is straightforward to integrate Eq. (\ref{Eq17}) using the method
described in \cite{Gundlach1}. The late-time evolution of a charged
scalar-field is independent of the form of the initial data. The
results presented here are of a Gaussian pulse on $u=0$
\begin{equation}\label{Eq19}
\psi (u=0,v)= A exp \left \{-\left [ (v-v_{0})/ \sigma \right ]^{2}
\right \}\  ,
\end{equation}
where the amplitude $A$ is physically irrelevant due to the linearity
of Eq. (\ref{Eq17}).
It should be noted that the evolution equation (\ref{Eq17}) is
invariant under the rescaling
\begin{equation}\label{Eq20}
r \to ar \  ,\  t \to at \  ,\  M \to aM \  ,\  Q \to aQ\  ,\  e \to e/a \  ,
\end{equation}
where $a$ is some positive constant.
The black-hole mass and charge are set equal to $M_{BH}=0.5$ 
and $Q_{BH}=0.45$, respectively.
We have chosen $e=0.01\ ,v_{0}=100$ and $\sigma =20$ (for both the
real and the imaginary parts of the charged field).
The numerical results for the $l$=0,1 and 2 modes are
shown in Fig. \ref{Fig3}. (from top to 
bottom, respectively).
This figure demonstrates the dependence of the late-time
tails at $i_{+}$ (here y=400) on the multipole index $l$.
A definite {\it power-law} fall off is manifest at late-times.
The numerical values of the
power-law exponents, describing the fall-off of the field at
{\it late} times are
$-1.97, -3.94$ and $-5.75$ for $l$=0,1 and 2, respectively. 
These values are to be compared with
the {\it analytically} predicted values of $-2.0, -4.0$
and $-6.0$, respectively.

In a second series of numerical investigation, we have studied
the evolution of a second perturbative charged scalar-field $\xi$
evolved on a {\it time-dependent} spacetime. This time-dependent
spacetime is determined
by the background solution $\phi$, while we ignore the contribution of
the field $\xi$ to the energy-momentum tensor (this is analogous to 
the case studied in \cite{Gundlach2} for a neutral field).
Resolving the field into spherical harmonics
$\xi=\sum\limits_{l,m} {\xi^{l}_{m}(t,r) Y_{l}^{m}(\theta,\varphi)}$
(and using the spherical symmetry of the time-dependent background)
one obtains a wave-equation for each multiple moment of the field $\xi$
\begin{equation}\label{Eq21}
{d \tilde \xi^{l}_{m} \over du}={1 \over {2r}}(g - \bar g)(\tilde \xi^{l}_{m}-
\xi^{l}_{m})-
{{Q^{2}} \over {2r^{3}}}(\tilde \xi^{l}_{m}- \xi^{l}_{m})g -
{ieQ \over 2r}g \xi^{l}_{m} -
ie \tilde \xi^{l}_{m} A_{u} -
{1 \over {2r}} l(l+1)g \xi^{l}_{m}\  ,
\end{equation}
[together with Eq. (\ref{Eq14})], where
\begin{equation}\label{Eq22}
\xi^{l}_{m} \equiv {1 \over r} \int_{0}^{r} \tilde \xi^{l}_{m} dr\  .
\end{equation}
The quantities $g, \bar g$ and $Q$ are still determined by the background
solution for the field $\phi$.
For the background field $\phi$ we choose the same initial form
as in Sec. \ref{Sec4}. The most interesting time-dependent backgrounds
are the {\it non-}collapsing ones. Thus, we take the {\it subcritical}
initial conditions $A$=0.002 and $e$=3 of Sec. \ref{Sec4}.
For the perturbation field $\xi$ we choose initial data of the
form Eq. (\ref{Eq16}) with $r_{0}=4.0$ and $\sigma =0.5$ for both 
the real and the imaginary parts of the field (again, the amplitude
of the perturbation field is physically irrelevant).
Figure 4 displays the time evolution of the perturbative charged scalar field 
$\xi$ (its real part) at a constant radius (here $r=10$) as a
function of Bondi time.
It is interesting that even for {\it time-dependent} spacetimes
the late-time behaviour of charged-fields is well described
by an inverse {\it power-law} fall-off.
The numerical values of the power-law exponents are
$-1.99$ for the $l$=0 mode (top panel) and $-4.16$ for the $l$=1
mode (bottom panel). These values should be compared with
the fixed background {\it analytically} predicted values 
of $-2.0$ and $-4.0$, respectively.

\section{Summary and physical implications}\label{Sec6}

We have studied the {\it nonlinear} gravitational
collapse of a massless {\it charged} scalar-field.
Following the predictions of the linearized {\it analytical}
theory (papers I and II) we have focused attention on the 
{\it asymptotic} late-time evolution of the charged-field.
Our main results and their physical implications are:

We have confirmed the existence of oscillatory 
inverse {\it power-law tails} in a {\it collapsing} spacetime 
along the asymptotic 
regions of future timelike infinity $i_{+}$, future null infinity $scri_{+}$ 
and along the future outer-horizon $H_{+}$.
The {\it nonlinear} dumping exponents are in excellent 
agreement with the {\it analytically} predicted ones.
In particular, we have verified the analytically conjectured 
dependence of the {\it charged} dumping exponents on the
dimensionless quantity $|eQ|$. This dependence on the spacetime
charge is contrasted to neutral perturbations, where the
dumping exponents are fixed integers which does {\it not} depend
on the spacetime parameters (namely, they are functions
of the multipole index $l$ only).
We have confirmed the existence of power-law tails even in
non-collapsing spacetimes (i.e. imploding and exploding shells).

We have also studied the late-time evolution
of non-spherical perturbations ({\it charged} test fields) on a fixed 
Reissner-Nordstr\"om background and on a time-dependent background.
In both cases we have found that the dumping exponents, describing
the fall-off of the charged field at late-times, are in good agreement
with the predictions of the linearized analytical theory (In
particular, we have verified the functional dependence of the 
dumping exponents on the multipole index $l$).

Our nonlinear results verify the analytic conjecture 
that a {\it charged} hair 
decays {\it slower} than a neutral one.
Furthermore, the existence of oscillatory inverse {\it power-law}
tails along the black-hole outer horizon 
suggests the occurrence of the
well-known phenomena of {\it mass-inflation} \cite{Poisson} 
along the Cauchy horizon of a {\it dynamically} formed charged
black-hole.

In a forthcoming paper we study 
the {\it fully nonlinear} gravitational
collapse of a charged scalar-field, using a different 
numerical scheme which is based on {\it double} null coordinates.
This scheme allows us to start with {\it regular}
initial conditions (at approximately past null infinity),
calculating the {\it formation} of the regular black-hole's
event horizon,
and continue the evolution all the way {\it inside} the black-hole.
Thus, this numerical scheme makes it possible to test the
mass-inflation conjecture during the gravitational 
collapse of a {\it charged} scalar-field.
To our knowledge, the mass-inflation scenario has {\it never}
been demonstrated explicitly before in {\it collapsing} situations 
(The numerical work of
Brady and Smith \cite{Brady} begins on a Reissner-Nordstr\"om
spacetime and the black-hole formation itself
was {\it not} calculated there).

\bigskip
\noindent
{\bf ACKNOWLEDGMENTS}
\bigskip

This research was supported by a grant from the Israel Science Foundation.

\begin{figure}
\centering
\noindent
\includegraphics[width=15cm]{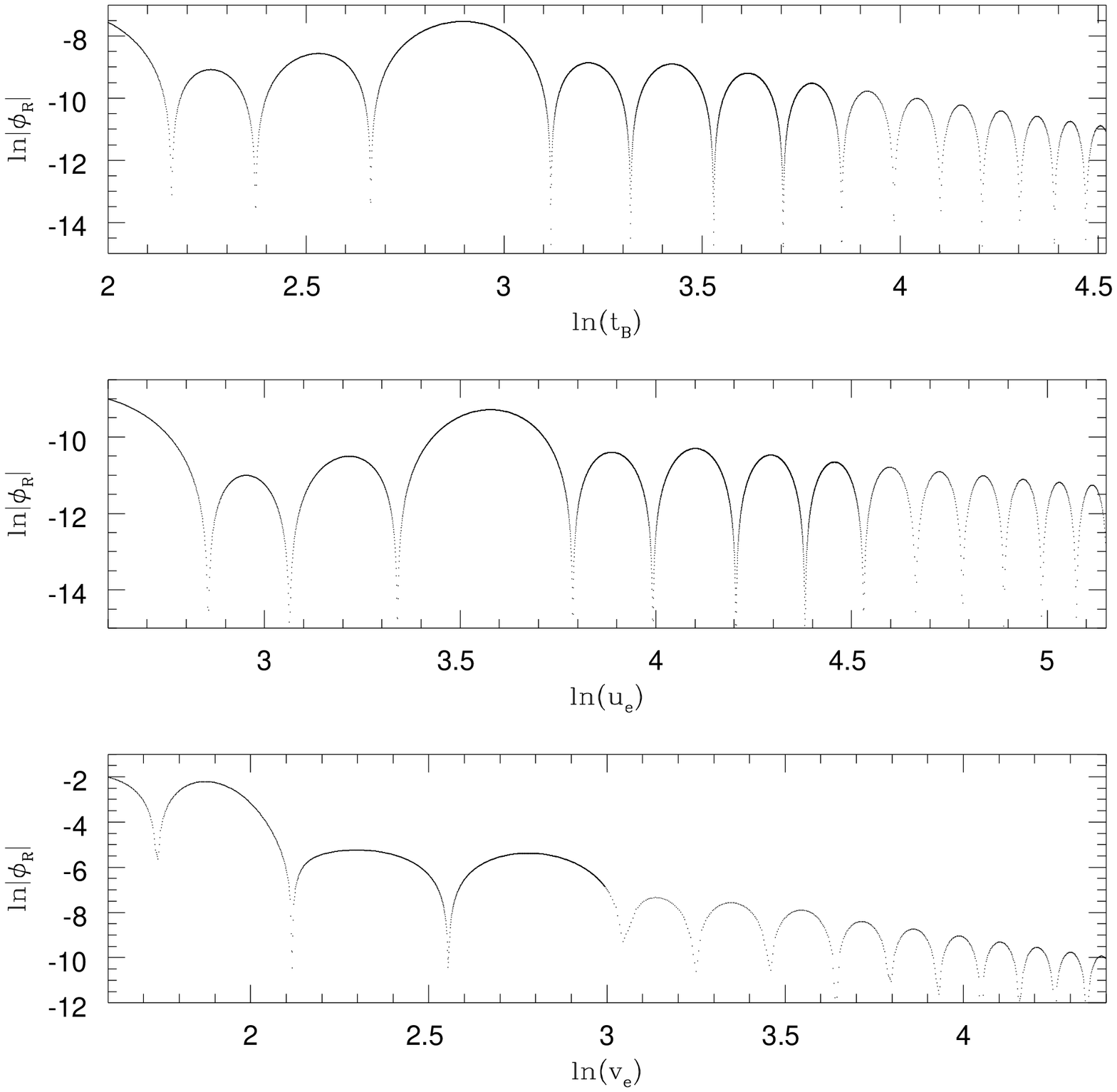}
\caption[Supercritical]{\label{Fig1}
Supercritical evolution of the charged field $|\phi_{R}|$ along the asymptotic
regions of timelike infinity (top panel), null infinity (middle
panel) and the black-hole outer horizon (bottom panel).
The initial data is a Gaussian distribution (multiplied by 
an $r^{2}$ factor).
The field's amplitude is $A$=0.005 and $e$=0.85.
The mass and charge of the formed black-hole are $M_{BH}=0.503$
and $Q_{BH}=-0.420$, respectively.
A definite oscillatory {\it power-law} fall-off is manifest
at late-times.
The {\it nonlinear} power-law exponents are $-1.78$ at timelike
infinity (here $r=10$), $-0.86$ along null infinity and $-1.96$ 
along the black-hole outer horizon.
These values are to be compared with the {\it analytically} 
predicted values of $-1.70, -0.85$ and $-1.70$, respectively.
The oscillation frequency of the charged field $\phi$ agrees with
the {\it analytically} predicted value to within $2\%$.} 
\end{figure}
 
\begin{figure}
\centering
\noindent
\includegraphics[width=15cm]{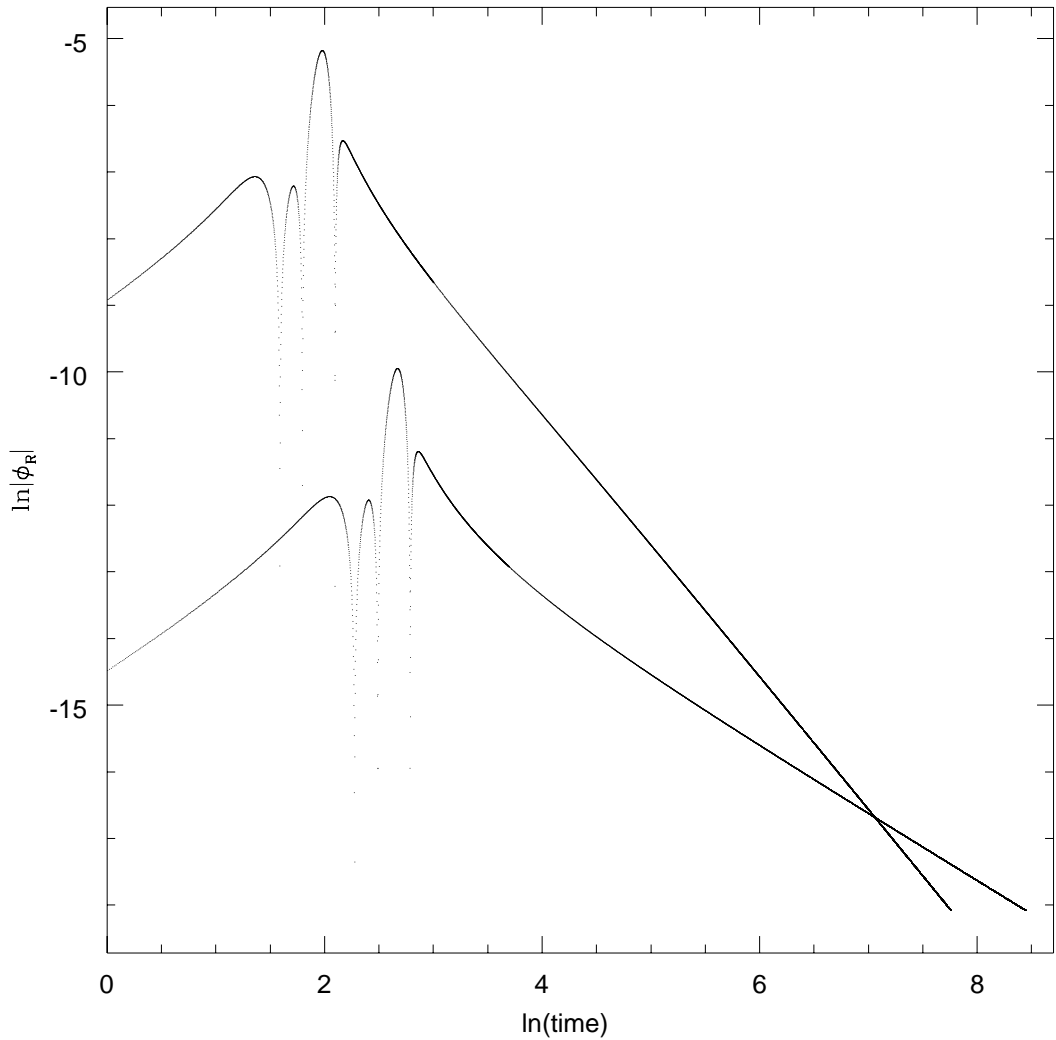}
\caption[Subcritical]{\label{Fig2}    
Subcritical evolution of the charged field $|\phi_{R}|$.
The initial form of the field is the same as in Fig. \ref{Fig1}.
The field's amplitude is $A$=0.002 and $e$=3. 
The field at future timelike infinity ($r=10$) is shown 
as a function of $t_{B}$. 
Along null infinity the field is shown as a function of $u_{e}$.
A definite {\it power-law} fall-off is manifest
at late-times.
The {\it nonlinear} power-law exponents are $-2.00$ at timelike
infinity and $-1.00$ along null infinity. 
These values are exactly the ones predicted by the linearized
{\it analytical} approach.}
\end{figure}

\begin{figure}
\centering
\noindent
\includegraphics[width=15cm]{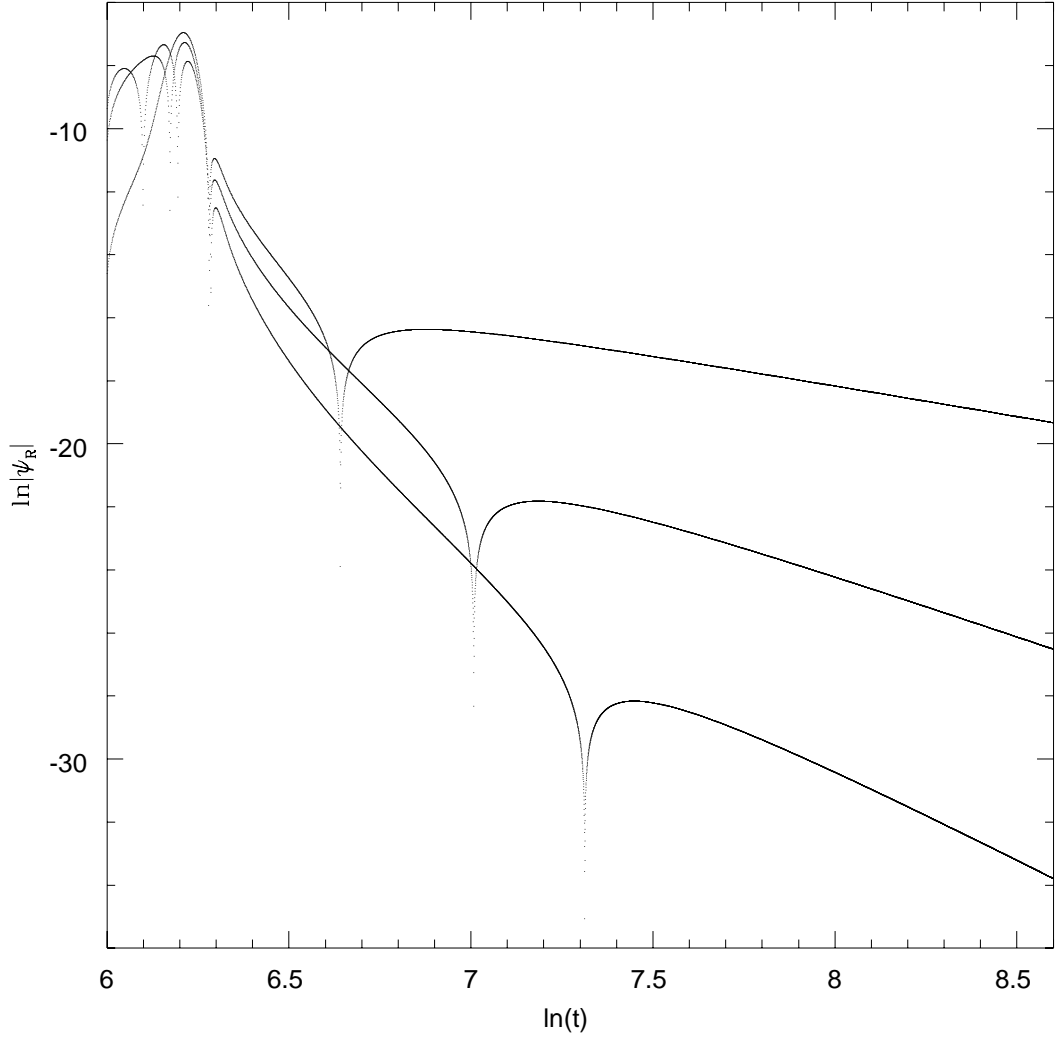}
\caption[various values of $l$]{\label{Fig3}
Evolution of the charged field $|\psi_{R} (y=400,t)|$
on a fixed Reissner-Nordstr\"om background for
different multipoles $l$=0,1 and 2 (from top to bottom, respectively).
The black-hole mass and charge are set equal to $M_{BH}=0.5$
and $Q_{BH}=0.45$, respectively. The field's charge is $e=0.01$.
The initial data is a Gaussian
distribution with $v_{0}=100$ and $\sigma =20$ (for both the
real and the imaginary parts of the field).
A definite {\it power-law} fall-off is manifest
at late-times.
The power-law exponents
are $-1.97, -3.94$ and $-5.75$ for the $l$=0,1 and 2 modes, respectively.
These values are to be compared with the {\it analytically}
predicted values of $-2.0, -4.0$ and $-6.0$.}
\end{figure}

\begin{figure}
\centering
\noindent
\includegraphics[width=15cm]{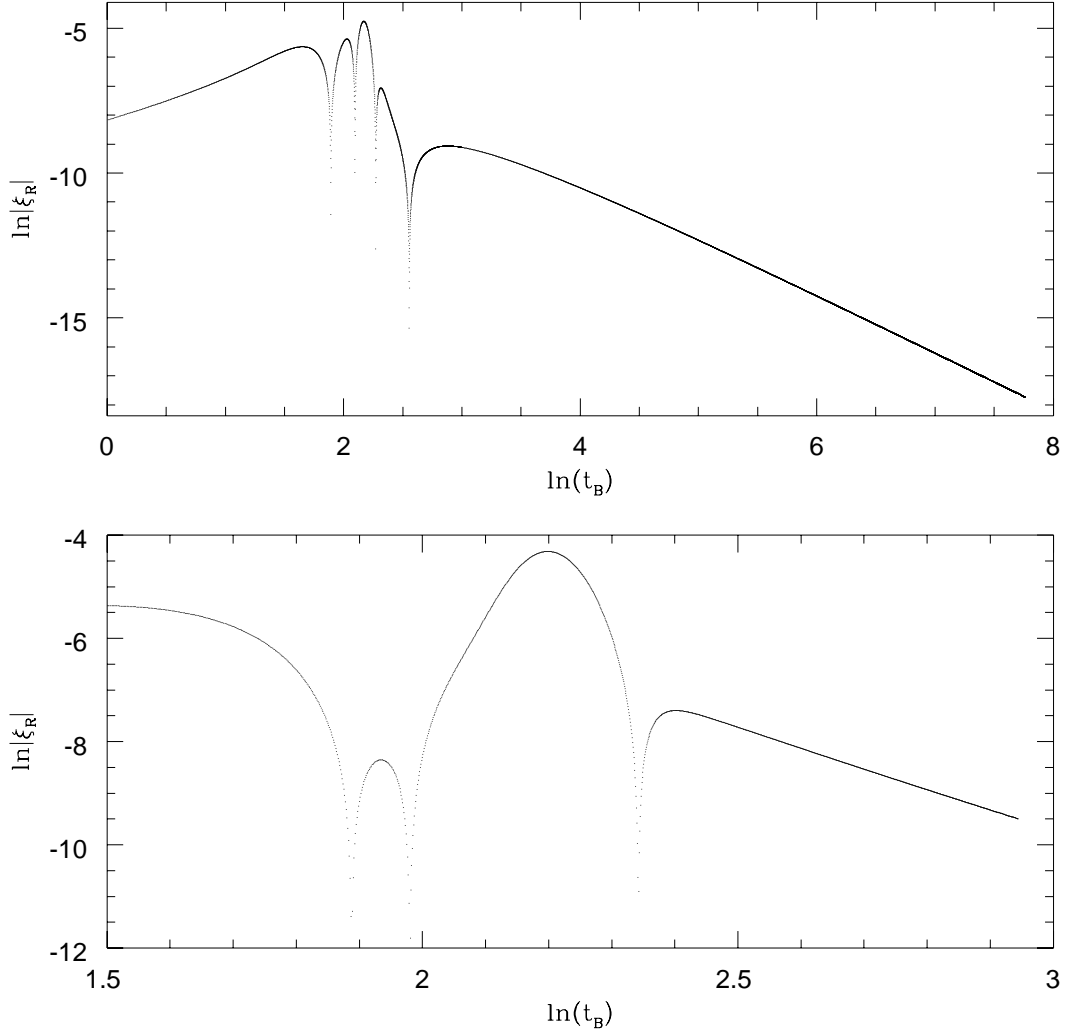}
\caption[time-dependent spacetime]{\label{Fig4}
Evolution of the charged test field $|\xi_{R} (r=10,t)|$ on a
{\it time-dependent} spacetime.
The initial data for the background field $\phi$ are those
of Fig. \ref{Fig2}. ({\it non-collapsing} case).
The initial data for the test field $\xi$ is
a Gaussian distribution (multiplied by 
an $r^{2}$ factor).
A definite {\it power-law} fall-off is manifest
at late-times.
The power-law exponents
are $-1.99$ for the $l$=0 mode (top panel) and $-4.16$ for 
the $l$=1 mode (bottom panel).
These values are to be compared with the {\it analytically}
predicted values of $-2.0$ and $-4.0$, respectively.}
\end{figure}

\begin{table}
\caption{Dependence of the dumping exponents at timelike infinity
$i_{+}$ on $eQ$}
\label{Tab1}
\begin{tabular}{cccc}
$e$ & $Q_{BH}$ & $2\beta+2$ & Nonlinear exponents\\
\tableline
0.50 & -0.241 & -1.97 & -1.99 \\
0.70 & -0.344 & -1.88 & -1.93 \\
0.85 & -0.420 & -1.70 & -1.78 \\
0.90 & -0.443 & -1.60 & -1.65 \\
\end{tabular}
\end{table}

\end{document}